\documentstyle[12pt]{article}
\title{`Natural Masslessness Conservation'
for neutrinos in two Higgs-doublet models}
\author{A.\ Barroso \\
\small Departamento de F\'{\i}sica, Faculdade de Ci\^{e}ncias,
Universidade de Lisboa \\
\small Campo Grande, C1, 1600 Lisboa, Portugal\\
   \\
and\\
   \\
Jo\~ao P.\ Silva\\
\small Department of Physics, Carnegie-Mellon University, \\
\small Pittsburgh, Pennsylvania 15213, U.S.A.}
\begin{document}
\maketitle
\begin{abstract}
We present a model which supplements the Standard Electroweak Model with
three right-handed neutrinos and one extra scalar doublet which does
not develop a vacuum expectation value. With the aid of a discrete symmetry
the neutrinos are kept strictly massless. This model has several interesting
features. It has unsuppressed lepton flavour violating processes,
in particular $\mu \rightarrow e \gamma$, hinting
at the possibility that these may soon be within experimental reach.
The $Z$ and $W$ interactions become non-diagonal at one loop level.
In particular, a non-trivial leptonic mixing matrix is seen to arise
from the clash between the charged gauge boson and the charged scalar
interactions.
\end{abstract}


\section{Introduction}

One of the interesting features of the leptonic sector is the fact that
neutrino masses are much smaller
than the other fermion masses.
To date, no experiment has unequivocally detected nonzero neutrino masses.
In the Standard Model (SM) \cite{gsw}
one precludes the existence of Dirac masses by not
having right handed neutrino fields in the theory.
This is a somewhat peculiar feature of the theory, since all other fermions
appear as both right handed (RH) and left handed (LH) fields.
Therefore we would like to look for a minimal extension of the SM which
does not have this shortcoming.
However, we must then face the problem of justifying
the smallness (or inexistence) of the neutrino Dirac mass terms.

This has been done in several extensions of the SM.
In some grand unified theories such as SO(10), a Majorana lepton number
violating mass term for the right handed neutrinos yields
neutrinos with small masses
through the see-saw mechanism \cite{seesaw}.
In several other models
neutrinos are massless and yet lepton
flavour is violated.
In some of these
\cite{nounit} \cite{valle}
this is due to the lack of unitarity
of the charged current mixing matrix which translates into non-universal
couplings for the neutral current interaction.
In others
\cite{susy},
R-parity violating terms in the superpotential
generate flavour violating neutral current interactions
with the electron and the u and d quarks.
Most of these models involve rather speculative assumptions about the
field content of the theory.

On the other hand, in the SM there is no justification
for the presence of only one Higgs doublet.
In fact considerable interest has arisen in multi-Higgs
doublet models,
ever since T.\ D.\ Lee showed \cite{spcpv} that one could
have spontaneous CP violation in the two-Higgs doublet model.
To avoid flavour changing neutral exchanges,
S.\ L.\ Glashow and S.\ Weinberg , and independently
E.\ Paschos, proposed
the introduction of a discrete symmetry
forcing each fermion to couple to only
one Higgs doublet : this is known as natural flavour conservation
\cite{nfc}.
In what follows we will use a similar device to show how one
can naturally suppress neutrino masses in a two Higgs doublet model with
three RH neutrino fields.

In section 2 we present our model.
In section 3  we study the experimental constraints arising from the
processes $l_2 \rightarrow l_1 \nu {\bar \nu}$,
$l_2 \rightarrow l_1 \gamma$ and
$l_3 \rightarrow l_2 l_1 {\bar l_1}$.
We also discuss briefly the decays $Z \rightarrow l_1 {\bar l_2}$ and
$W \rightarrow l_1 {\bar \nu_2}$.
In the last section we
draw our conclusions. The appendix contains
several derivations
needed to establish the experimental bounds on the theory.

\section{The Model}

Our model has, in addition to the SM fields,
three right handed (RH) neutrino fields
\footnote{We could equally well construct a theory with any other
number of right handed neutrinos; for example, one or two. We chose
three since we seek a `symmetric'
theory with as many right handed singlet fermions as
there are left handed fermions.},
$N_R$, and one extra
Higgs doublet, $H_2$. The Higgs potential is chosen so that
$H_2$ does not get a vacuum expectation value. In addition,
a discrete symmetry is introduced under which
\begin{equation}
N_R \rightarrow - N_R\ , \hspace{5mm} H_2 \rightarrow - H_2\ ,
\end{equation}
and all other fields remain invariant. The Yukawa and weak lagrangeans are,
\begin{eqnarray}
- {\cal L}_Y    & = &
{\bar L}_L H_1 \Gamma_1 C_R +
{\bar L}_L (i \sigma_2 H_2^\ast) \Gamma_2 N_R\ + h.c. ,\\
{\cal L}_W      & = &
\frac{g}{\sqrt{2}} {\bar N}_L \gamma^\mu C_L W_\mu^+ + h.c.\ ,
\end{eqnarray}
where ${\bar L}_L$
is the left handed (LH) lepton doublet, $N_R$ ($C_R$) is the RH
neutrino (charged lepton) singlet and $\Gamma_1$, $\Gamma_2$ are
$3 \times 3$ Yukawa coupling matrices.
For later use the Higgs fields will be written as
\begin{equation}
H_1
= \left( \begin{array} {c}
                G^+ \\
                \frac{1}{\sqrt{2}} (v + H^0 + i G^0)
             \end{array} \right)\ ,
\hspace{10mm}
H_2
= \left( \begin{array} {c}
                H^+ \\
                \frac{1}{\sqrt{2}} (R + i I)
             \end{array} \right)\ ,
\end{equation}

We can diagonalize the charged lepton mass matrix while
keeping the charged current diagonal with transformations,
\begin{eqnarray}
{\bar L}_L \equiv ({\bar N}_L , {\bar C}_L)
& = &
({\bar \nu}_L , {\bar l}_L) U_L^\dagger \ ,\\
C_R
& = &
U_{C_R} l_R\ ,
\end{eqnarray}
where
\begin{equation}
D_l \equiv diag(m_e,m_\mu,m_\tau) =
\frac{v}{\sqrt{2}} U_L^\dagger \Gamma_1 U_{C_R}\ .
\end{equation}
We can still use the fact that $\Gamma_2$ can be diagonalized by a bi-unitary
transformation, $\Gamma_2 = X N_\nu U_{N_R}^\dagger =
X diag(n_1,n_2,n_3) U_{N_R}^\dagger$, and the freedom to redefine the RH
neutrino fields by,
\begin{equation}
N_R = U_{N_R}^\dagger \nu_R\ ,
\end{equation}
to write,
\begin{eqnarray}
- {\cal L}_Y
& = &
... + (1+H^0/v)  {\bar l}_L D_l l_R - H^- {\bar l}_L M^\dagger \nu_R
- H^+ {\bar \nu}_R M l_L \nonumber\\
&  &
+R/\sqrt{2} [{\bar \nu}_R M \nu_L + {\bar \nu}_L M^\dagger \nu_R]
+i I/\sqrt{2} [{\bar \nu}_R M \nu_L - {\bar \nu}_L M^\dagger \nu_R]\ ,\\
{\cal L}_W
& = &
\frac{g}{\sqrt{2}} {\bar N}_L \gamma^\mu C_L W_\mu^+ + h.c.\ .
\end{eqnarray}
Note that all fields in $H_2$ involve the same coupling
$M^\dagger = B^\dagger N_\nu$: the charged scalars link $\nu_R$ and
$l_L$; the neutral scalars link $\nu_R$ and $\nu_L$.

A simple example of lepton flavour violations at low energies occurs
in the decay of a LH lepton of flavour i into a LH lepton of flavour j
and some gauge boson. This arises through a one loop
diagram with intermediate $\nu_R$ and $H_2$ fields
(Cf., for example, fig.\ 1),
and will be proportional to,
\begin{equation}
\Omega_{ij} =
\sum_{k=1}^{3} M_{ki} M^\dagger_{jk} =
\sum_{k=1}^{3} |n_k|^2 B_{ki} B_{kj}^\ast\ ,
\label{eq:master}
\end{equation}
showing a suppression in the limit in which the $n_K$ values are
close to each other.

In our model, overall
lepton number conservation is imposed
forbidding Majorana mass terms and we shall
assume that there is no CP violation in the lepton sector (that is,
the matrix $N_\nu$ is real and the matrix $B$ is orthogonal; but we
will keep our formulas general).
Moreover, we shall take the new scalar masses to be within
an order of magnitude, or so,
of the electroweak scale, $v=246 GeV$.

\section{Experimental Bounds}

The first important limit on our theory comes from measurement of
$G_F = 1/\sqrt{2}v$ in the muon decay as compared to that made
in the quark sector. For untagged neutrino flavours,
we have both charged
W-exchange and charged H-exchange tree level diagrams. Due to their
different chiral structures, these do not interfere and
one can easily find,
\begin{equation}
\Gamma(\mu \rightarrow e \nu {\bar \nu}) =
\frac{G_F^2 m_\mu^5}{192 \pi^2} (1+\frac{v^4}{16 M_H^4}
\Omega_{\mu \mu} \Omega_{e e})\ ,
\label{eq:Gmu}
\end{equation}
where $M_H$ is the mass of the charged scalars H.

In our model, the quark mixing matrix is unitary and, therefore,
the sum of the magnitudes squared
of the elements in its first row must add to one.
Any deviation will be a measure of how much the expression within
parenthesis
in Eq.~\ref{eq:Gmu} deviates from one. Using the result \cite{pdb},
\begin{equation}
|V_{ud}|^2+|V_{us}|^2+|V_{ub}|^2 = 0.9977 \pm .0030
\end{equation}
we find
\footnote{The radiative corrections are dominated by QED effects and
are therefore down by $\alpha/4 \pi$; almost an order of
magnitude bellow our bound on new physics. However, to improve
these constraints, such as by measuring the neutrino spectra in muon
decay, these QED radiative corrections must be taken into account.
An excellent discussion of this can be found in reference \cite{GWF}.}
,
\begin{equation}
\frac{v^4}{M_H^4} \Omega_{\mu \mu} \Omega_{e e} < 0.0848\ .
\end{equation}
Noting from Eq.~\ref{eq:master} that $\Omega_{l l} \geq  0$,
we find that, at tree level, the present sign of the deviation from the mean
is consistent with this model.
The effect of the one loop amplitudes will be briefly discussed at
the end of this section.
In any event, it is a good approximation
to take the value of $G_F$ as measured from muon decay,
in the following analysis.

The situation concerning the tau branching ratios, is still
unclear \cite{hayes}.
This is particularly so for the decay into electron and two neutrinos
where the relation
\begin{equation}
\tau_\tau = \tau_\mu
(\frac{G_\mu}{G_\tau})^2 (\frac{m_\mu}{m_\tau})^5
B(\tau \rightarrow e \nu {\bar \nu})\ ,
\end{equation}
has only recently become consistent with the Standard Model
($G_\mu = G_\tau$) \cite{erratum}.
It is widely believed that this problem
will continue to disappear with further experiments.
Therefore, we shall not derive any constraints from
this decay.

As can be seen from Eq.~\ref{eq:master}, we will have lepton
flavour violation in processes such as $l_2 \rightarrow l_1 \gamma$
or $l_3 \rightarrow l_1 l_1 l_2$. The first of these processes
has one-loop contributions from the diagrams in fig. 1, involving
an intermediate charged scalar and RH neutrinos.
Clearly no such diagram exists with an intermediate gauge boson.

The most general form of the invariant amplitude for the process
in fig.1 is,
\begin{eqnarray}
{\cal M}
& = &
\epsilon_\mu T_\gamma^\mu \nonumber\\
   & = &
\epsilon_\mu {\bar u}_1(p_1)
[ i m_2 \sigma^{\mu \nu} k_\nu (A_\gamma + B_\gamma \gamma_5)
\nonumber\\
   &   & \hspace{30mm}
+ k^2 \gamma^\mu (C_\gamma- D_\gamma \gamma_5)
- k^\mu (E_\gamma + F_\gamma \gamma_5)] u_2(p)\ ,
\end{eqnarray}
where $k=p-p_1$, gauge invariance guarantees that,
\begin{equation}
C_\gamma= \frac{E_\gamma}{m_2 - m_1}\ ,
\hspace{10mm} D_\gamma= \frac{F_\gamma}{m_2 + m_1}\ ,
\label{eq:help}
\end{equation}
and the factor of $k^2$ in the $\gamma^\mu$ terms reflects the fact
that this contribution vanishes in the $k^2=0$ limit.

The diagrams in figs.\ 1b and 1c only contribute
to the $\gamma^\mu$ terms.
So, it is easier to use the diagram in fig.\ 1a to find $A_\gamma$,
$B_\gamma$,
$E_\gamma$ and $F_\gamma$ and then use Eq.~\ref{eq:help} to get
$C_\gamma$ and $D_\gamma$.
We relegate the full expressions to the appendix.
To lowest order in $k^2$, and using
$M_H^2 >> m_2^2 >> m_1^2$, we find
\begin{eqnarray}
A_\gamma \approx B_\gamma
& \approx &
\frac{e \Omega_{21}}{32 \pi^2} \frac{1}{12 M_H^2}\ ,
\label{eq:Al}\\
C_\gamma \approx D_\gamma
& \approx &
\frac{e \Omega_{21}}{32 \pi^2} \frac{1}{18 M_H^2}\ .
\label{eq:Cl}
\end{eqnarray}

For the physical process $l_2 \rightarrow l_1 \gamma$, the photon
is on mass shell and only the $\sigma^{\mu \nu}$ terms contribute,
\begin{equation}
\Gamma(l_2 \rightarrow l_1 \gamma) = \frac{m_\mu^5}{8 \pi}
(|A_\gamma|^2 + |B_\gamma|^2)\ .
\end{equation}
In particular, for the muon decay we can use,
\begin{equation}
\Gamma(\mu \rightarrow e {\bar \nu}_e \nu_\mu)
\approx \frac{G_F^2 m_\mu^5}{192 \pi^3}
\approx \Gamma(\mu \rightarrow all)\ .
\end{equation}
As pointed out above,
the first sign is not a strict equality since, in this model,
there is also  an intermediate charged scalar diagram.
Thus,
\begin{equation}
B(\mu \rightarrow e \gamma) = \frac{\alpha}{32 \pi} \frac{1}{12}
|\sum_{k=1}^{3} (\frac{v n_k}{M_H})^2 B_{k \mu} B_{k e}^\ast|^2\ ,
\end{equation}
and the model is already constrained by experiment.
This should be compared
with what would happen if only RH neutrinos
and Dirac masses $m_k$ were added to the SM. In that case,
to lowest order in an expansion in powers of
$m_k/M_W$ one obtains \cite{chengli},
\begin{equation}
B(\mu \rightarrow e \gamma) = \frac{\alpha}{32 \pi} 3\
|\sum_{k=1}^{3} (\frac{m_k}{M_W})^2 U_{k \mu} U_{k e}^\ast|^2\ ,
\end{equation}
where U is the lepton mixing matrix \cite{bp} and there is no term
of order zero due to the GIM mechanism \cite{gim}.
Therefore, even if the neutrino masses saturate the cosmological bound,
we get a branching ratio of the order of $10^{-40}$ !
Furthermore, even if one includes both Dirac and Majorana masses,
only considerable fine tuning would lead to experimentally relevant
values for this branching ratio \cite{chengli2}.
By contrast, our model is already constrained by experiment.

{}From the experimental result \cite{pdb} of
$B(\mu \rightarrow e \gamma) < 4.9 \times 10^{-11}$
we find,
\begin{equation}
\frac{v^2}{M_H^2} |\Omega_{\mu e}| < 2.8 \times 10^{-3}\ .
\end{equation}
It seems natural to suspect that the $n_k$ aren't much smaller than one,
$M_H$ is within an order of magnitude of $v$ and that the
$B$ matrix might have some off-diagonal elements of
order $0.1$, say. This would raise the exciting prospect of a positive
result with further experiments.
For the tau decays,
$B(\tau \rightarrow \mu \gamma) < 5.5 \times 10^{-4}$
and $B(\tau \rightarrow e \gamma) < 2.0 \times 10^{-4}$
lead to the considerably weaker bounds,
\begin{equation}
\frac{v^2}{M_H^2} |\Omega_{\tau \mu}| < 22\ ,
\hspace{10mm}
\frac{v^2}{M_H^2} |\Omega_{\tau e}| < 13\ .
\end{equation}

Similarly, we will have flavour violation in processes such as
$l_2 \rightarrow 3 l_1$ due to the diagrams in fig.\ 2.
Clearly, the Z diagram is very suppressed due to the $Z$ propagator
and we can ignore it.
The general
expressions have been derived in the appendix. For our model,
\begin{eqnarray}
\Gamma(l_2 \rightarrow 3 l_1)
& = &
\frac{G_F^2 m_2^5}{192 \pi^3} \frac{1}{(32 \pi^2)^2}
\frac{v^4}{M_H^4} |\Omega_{21}|^2 \ \
[\frac{8 \pi^2 \alpha^2}{9} (2 \ln{(\frac{m_2}{2 m_1})}-\frac{23}{12})
\nonumber\\
   &   &
\hspace{44mm}
+\frac{1}{8}|\Omega_{11}|^2+\frac{4 \pi \alpha}{9} Re\{\Omega_{11}\}]\ ,
\end{eqnarray}
and, using the experimental constraints we find,
\begin{eqnarray}
\frac{v^2}{M_H^2} |\Omega_{\mu e}|
\sqrt{1+36.4 |\Omega_{e e}|^2 +2.96 Re\{\Omega_{e e}\}}
& \leq &
5.4 \times 10^{-3}\ ,
\nonumber\\
\frac{v^2}{M_H^2} |\Omega_{\tau \mu}|
\sqrt{1+114 |\Omega_{\mu \mu}|^2 +9.28 Re\{\Omega_{\mu \mu}\}}
& \leq & 90\ ,
\nonumber\\
\frac{v^2}{M_H^2} |\Omega_{\tau e}|
\sqrt{1+20.6 |\Omega_{e e}|^2 +1.68 Re\{\Omega_{e e}\}}
& \leq &
48\ .
\end{eqnarray}
These are limits on different quantities than the ones showing up
above, but point to roughly the same order of magnitude.

In this model, one also has lepton flavour violating $Z$ decays but
these have far worse experimental constraints than the flavour
violating lepton decays.
In addition, the theory already predicts that, for example,
$B(Z\rightarrow \mu^+ e^-)$ should be much smaller than
$B(\mu \rightarrow e \gamma)$. This arises from the
fact that the dominating $W$ exchange decay mode
of the muon has a three body final state phase space suppression,
together with the fact that the $Z$ has many decay modes.

Using the results derived in the appendix we find that since
$M_Z^2 >> m_2^2 >> m_1^2$, the dominant contribution comes from the
$\gamma^\mu$ terms yielding,
\begin{equation}
\Gamma(Z \rightarrow l_1 {\bar l_2}) =
\frac{M_Z}{12 \pi} [|C_Z(k^2=M_Z^2)|^2 + |D_Z(k^2=M_Z^2)|^2]\ ,
\end{equation}
where, for $M_H^2 >> M_Z^2$,
\begin{equation}
D_Z(k^2=M_Z^2)
\approx C_Z (k^2=M_Z^2)
\approx
\frac{e \Omega_{21}}{32 \pi^2}\ \cot(2 \theta_W)\
\frac{M_Z^2}{18 M_H^2}\ .
\end{equation}
{}From this one can easily derive a
relation between the two branching ratios, namely,
\begin{equation}
B(Z \rightarrow e^- \mu^+) \approx
3.8 \times 10^{-5} B(\mu^- \rightarrow e^- \gamma)\ ,
\end{equation}
showing that even for the tau (for which the coefficient  is around
$5.6$ times larger) this rate is unmeasurably small.

A similar analysis may be performed for the charged gauge interactions.
The results can again be found in the appendix. The most important feature
is the appearance at one loop level of off-diagonal
vertices with $W$, leptons and neutrinos.
In particular, the left-handed vector interaction acquires a
non-diagonal leptonic mixing matrix,
\begin{equation}
V_{12} = \frac{\Omega_{21}}{32 \pi^2} f[M_R,M_I,M_H]\ ,
\end{equation}
where the function
\begin{equation}
f[M_R,M_I,M_H] = [\frac{1}{2} \frac{M_H^2+M_R^2}{M_H^2-M_R^2}
\ln{(M_R/M_H)}]+ [R \rightarrow I] + 1
\end{equation}
vanishes for $M_R=M_I=M_H$, is negative elsewhere and remains of order
$(-)1$ for any reasonable ratios of scalar masses. Note that this mixing
matrix is not unitary and therefore cannot be absorbed by a
redefinition of the left-handed neutrinos.
Thus we find the
interesting feature of a non trivial leptonic mixing matrix even with
massless neutrinos \cite{nounit} with its possible implications for neutrino
propagation in matter \cite{valle}.

To be exhaustive we should now go back and reassess our calculation for the
muon decay. Indeed, the one loop corrections to the W-exchange amplitude
might be comparable to the H-exchange amplitude if
\begin{equation}
\frac{\Omega_{\mu e}}{32 \pi}\   f[M_R,M_I,M_H]
\approx
\frac{v^2}{4 M_H^2} \sqrt{\Omega_{\mu \mu} \Omega{e e}}
\end{equation}
This will depend on the masses of the new scalars and on the $\Omega_{i j}$.
It is also clear that, for example, the analysis of Beta decay must
take this into account modifying the extraction of the CKM matrix elements.

\section{Conclusions}

We have developed a model with massless neutrinos inspired by a
minimal `democracy' assumption:
there should exist a right-handed singlet partner for every left-handed
particle; the fundamental scalars might also exist in several families.
This was achieved at the expense of a `discriminatory' discrete symmetry.

This model has several interesting characteristics.
Contrary to what happens if one adds massive Dirac neutrinos
to the SM, in which case the cosmological limit imposes
minute lepton flavour
violations,
in this model, such lepton flavour violating processes
might be within experimental reach, especially
$\mu \rightarrow e \gamma$.
Although one also gets non-diagonal $Z$
decays we found that these have levels beyond experimental verification.
Finally, one also finds non-diagonal $W$ interactions, and this despite
the fact that the neutrinos are massless. A novel feature is that this comes
about as the result of a clash between the $W$ and the $H$ interactions
with leptons.

This new sector of the theory will also have implications for
Cosmology. In particular, the new interactions must be weak enough
to decouple the new particles early enough not to affect
significantly primordial nucleosynthesis. This work
is currently under way.

\appendix

\section{Derivation of the invariant amplitudes}

In this appendix we derive the expressions for the
$l_2 \rightarrow l_1 \gamma$,
$l_2 \rightarrow l_1 Z$ and
$l_3 \rightarrow l_1 {\bar l_1} l_2$ decays, in our model.
The invariant amplitude for $l_2 \rightarrow l_1 \gamma$,
${\cal M} = \epsilon_\mu T_\gamma^\mu$ with
\begin{equation}
T_\gamma^\mu =
{\bar u}_1(p_1)
[ i m_2 \sigma^{\mu \nu} k_\nu (A_\gamma + B_\gamma \gamma_5)
+ k^2 \gamma^\mu (C_\gamma- D_\gamma \gamma_5) -
k^\mu (E_\gamma + F_\gamma \gamma_5)] u_2(p)\ ,
\end{equation}
may be found by computing solely the Feynman diagram of fig.~1a.
This is due to the fact that figs. 1b and 1c only contribute to the
vector and axial vector parameters which are easier to find with the
help of the Ward identity.
We find
\begin{eqnarray}
A_\gamma(k^2)
& = &
\frac{e \Omega_{21}}{32 \pi^2 m_2}
\int_{0}^{1} dx_1 \int_{0}^{x_1} dx_2
\frac{m_2(x_1-x_2)+m_1(1-x_1)}{\Delta(x_1,x_2)} (x_2)\ ,
\label{eq:A}\\
E_\gamma(k^2)
& = &
\frac{e \Omega_{21}}{32 \pi^2}
\int_{0}^{1} dx_1 \int_{0}^{x_1} dx_2
\frac{m_2(x_1-x_2)+m_1(1-x_1)}{\Delta(x_1,x_2)} (2 x_1-1-x_2)\ ,
\label{eq:E}
\end{eqnarray}
where,
\begin{equation}
\Delta(x_1,x_2) =
M_H^2 (1-x_2)-m_2^2(x_1-x_2)x_2-m_1^2(1-x_1)x_2 -k^2(x_1-x_2)(1-x_1)\ .
\end{equation}
The parameter $B_\gamma$ ($F_\gamma$) has the same expression as
$A_\gamma$ ($E_\gamma$), except
for a minus sign for the $m_1$ term in Eq.~\ref{eq:A}
(Eq.~\ref{eq:E}). Taking $M_H^2>>m_2^2>>m_1^2$, we find the result in
Eq.~\ref{eq:Al}, to lowest order in $k^2$.
The parameters $E_\gamma$ and $F_\gamma$ do not enter in either process
but, as mentioned above, we can use them to find $C_\gamma$ and
$D_\gamma$ through Eq.~\ref{eq:help}. The result,
with the same approximation, is in Eq.~\ref{eq:Cl}.

The invariant amplitude for $l_2 \rightarrow l_1 Z$ has the
same structure,
\begin{equation}
T_Z^\mu =
{\bar u}_1(p_1)
[ i m_2 \sigma^{\mu \nu} k_\nu (A_Z + B_Z \gamma_5)
+ \gamma^\mu (C_Z- D_Z \gamma_5) -
k^\mu (E_Z + F_Z \gamma_5)] u_2(p)\ ,
\end{equation}
except that, due to the mass of the $Z$, the $\gamma_\mu$ terms do not
vanish in the $k^2=0$ limit.
The calculation is just a repetition of that for the photon. The
difference in the diagram of fig.\ 1a (and therefore in the
$\sigma^{\mu \nu}$ and $k^\mu$ terms) is just due to the
$\xi_Z = \cot{(2 \theta_W)}$ ratio between the $HHZ$ and
$HH\gamma$ vertices. This is also the relevant ratio for the
$\gamma^\mu \gamma_L$ term of figs.\ 1b and 1c while the
$\gamma^\mu \gamma_R$ term involves $-\tan{\theta_W}$ (and
will be proportional to $m_1 m_2$ since one chirality flip is
needed in each external leg to reproduce this structure).
It is then easy to see that $\alpha_Z = \xi_Z \alpha_\gamma$
for $\alpha=A,B,E,F$ while,
\begin{eqnarray}
C_Z (k^2) & = &
\xi_Z [\eta_Z C_0 + k^2 C_\gamma(k^2)]\ ,
\nonumber\\
D_Z (k^2) & = &
\xi_Z [-\eta_Z C_0 + k^2 D_\gamma(k^2)]\ ,
\end{eqnarray}
where,
\begin{equation}
C_0 = \frac{e \Omega_{21}}{32 \pi^2} \frac{m_1 m_2}{m_2^2-m_1^2}
\int_{0}^{1} t dt \ln{[\frac{M_H^2 + m_1^2 (t-1)}{M_H^2 + m_2^2 (t-1)}]}\ ,
\end{equation}
and,
\begin{equation}
\eta_Z = \tan{(\theta_W)} \tan{(2 \theta_W)} + 1\ .
\end{equation}
In the same limit, $M_H^2>>m_2^2>>m_1^2$, we get
\begin{equation}
C_0 \approx \frac{e \Omega_{21}}{32 \pi^2} \frac{m_1 m_2}{6 M_H^2}\ .
\end{equation}

A straightforward calculation leads to
\begin{equation}
\Gamma(Z \rightarrow l_1 {\bar l_2}) =
\frac{M_Z}{12 \pi} [(|C_Z|^2+|D_Z|^2)
- 3 m_2^2 Re\{A_Z C_Z^\ast + B_Z D_Z^\ast\}
+m_2^2 M_Z^2/2 (|A_Z|^2+|B_Z|^2)]\ ,
\end{equation}
with the parameters evaluated at $k^2=M_Z^2$. For $M_H^2>>M_Z^2$ we
can use the approximate expressions derived for the photon parameters
in lowest order of $k^2/M_H^2$ to  get,
\begin{eqnarray}
C_Z(k^2=M_Z^2)& = &
\frac{e \Omega_{21}}{32 \pi^2} \xi_Z
\{ \frac{\eta_Z}{6} \frac{m_1 m_2}{M_H^2} + \frac{1}{18} \frac{M_Z^2}{M_H^2}
+ O (M_Z/M_H)^4 \}
\nonumber\\
   & \approx &
\frac{e \Omega_{21}}{32 \pi^2} \xi_Z \frac{M_Z^2}{18 M_H^2}\ ,
\end{eqnarray}
and similarly for $D_Z$.

A similar calculation shows that the W interactions also become off diagonal
at one loop level.
For the $Z$ and $\gamma$ interactions, the off-diagonal couplings are
excluded by symmetry at tree level.
Therefore, they are protected
against one loop divergences.
For the $W$ interactions the situation is different since
the lack of such off-diagonal couplings corresponds to a basis choice
and is not dictated by any symmetry.
In this case, the infinities are cancelled by the contributions from
the counterterms, which we calculated in the on-mass-shell
renormalization scheme \cite{soares}.
For the process $l_{2_L} \rightarrow W^- \nu_{1_L}$,
we find,
\begin{equation}
T_W^\mu =
{\bar u}_{\nu_1}(p_1)
[ i m_2 \sigma^{\mu \nu} k_\nu A_W (1 + \gamma_5)
+ \gamma^\mu C_W (1 - \gamma_5) -
k^\mu E_W (1 + \gamma_5)] u_2(p)\ ,
\end{equation}
where, in the limit that the square of the new scalar masses
($M_H^2$ for the charged scalars,
$M_R^2$ for the neutral scalar and $M_I^2$ for the pseudoscalar) are much
larger than $M_W^2$, and to first order in $k^2$, we get,
\begin{eqnarray}
A_W & = &
\frac{g}{2 \sqrt{2}} \frac{\Omega_{21}}{32 \pi^2}
\{
[\frac{1}{6(M_R^2-M_H^2)^2}
(M_H^2 - M_R^2 + M_R^2 \ln{(M_R^2/M_H^2)})] + [R \rightarrow I] \}\ ,
\\
C_W & = &
\frac{g}{2 \sqrt{2}} \frac{\Omega_{21}}{32 \pi^2} \{
[\frac{1}{2} \frac{M_H^2+M_R^2}{M_H^2-M_R^2} \ln{(M_R/M_H)}]+
[R \rightarrow I] + 1\} \ ,
\\
E_W & = &
\frac{g}{2 \sqrt{2}} \frac{\Omega_{21}}{32 \pi^2} m_2 \{
[\frac{M_R^2}{3(M_R^2-M_H^2)^3} (\ 2(M_H^2-M_R^2)+
(M_H^2+M_R^2) \ln{(M_R^2/M_H^2)}\ )]
\nonumber\\
   &  &
+ [R \rightarrow I]
\} \ .
\end{eqnarray}
As an easy check, we note that, except for a factor of $g/\sqrt{2}$ instead
of $e$, we recover the results for $l_2 \rightarrow l_1 \gamma$ when
$M_R^2 = M_I^2 = M_H^2$. In particular, in that limit,
we obtain $C_W=0$ reflecting
the fact that the vector coefficients for $l_2 \rightarrow l_1 \gamma$
vanish at zero momentum transfer.

For the process
$l_2 \rightarrow l_1 {\bar l_1} l_1$,
we get contributions from fig.\ 2. Note that there is no WW-box
contribution since, at tree level, there is no mixing matrix
in the lepton sector.
Similarly, chirality considerations exclude the WH-box diagram.

The amplitude for this process can be written as,
\begin{equation}
{\cal M}(l_2 \rightarrow l_1 {\bar l_1} l_1)=
{\cal M}(p_1,p_2) - {\cal M}(p_2,p_1)
\end{equation}
with,
\begin{equation}
{\cal M}(p_1,p_2)=
{\cal M}^\gamma(p_1,p_2) + {\cal M}^H(p_1,p_2) +
{\cal M}^Z(p_1,p_2)
\end{equation}
where ${\cal M}^\gamma$ is the one photon exchange amplitude and
${\cal M}^H$ is the HH-box contribution.
${\cal M}^Z$ is the $Z$-exchange amplitude which is very suppressed by the
$Z$ propagator since the momentum transfer is bound by $m_2^2$.
Therefore we will ignore this contribution.
Using the expression found above, it is easy to get,
\begin{equation}
{\cal M}^\gamma(p_1,p_2) = [T_\gamma^\mu]
[{\bar u}_1(p_2) e \gamma_\mu v_1(p_3)]
/k_1^2\ ,
\end{equation}
with $k_1=p-p_1$.
Similarly,
\begin{equation}
{\cal M}^H(p_1,p_2) = S
[{\bar u}_1(p_1) \gamma^\mu (1-\gamma_5)u_2(p)]
[{\bar u}_1(p_2) \gamma_\mu (1-\gamma_5)v_1(p_3)]\ ,
\end{equation}
where, in our model,
\begin{equation}
S= - \frac{\Omega_{21} \Omega_{11}}{256 \pi^2 M_H^2}\ .
\end{equation}
In the limit $m_2 >> m_1$, this leads to a decay rate,
\begin{equation}
\Gamma(l_2 \rightarrow 3 l_1) =
\frac{m_2^5}{768 \pi^3}
(\Gamma^\gamma+\Gamma^H+\Gamma^{\gamma H})\ ,
\end{equation}
where
\begin{eqnarray}
\Gamma^\gamma
& = &
4[4 \ln{(\frac{m_2}{2 m_1})} -\frac{13}{6}](|A|^2+|B|^2)
-12 Re\{A C^\ast+B D^\ast\} +3(|C|^2+|D|^2)\ ,
\nonumber\\
\Gamma^H
& = &
16 |S|^2\ ,
\nonumber\\
\Gamma^{\gamma H}
& = &
8 Re\{(C+D-2A-2B)S^\ast)\}\ ,
\end{eqnarray}
and the subscript $\gamma$ is implied.

A process such as $l_3 \rightarrow l_1 {\bar l_1} l_2$ has a
similar expression in the limit $m_3 >>m_2, m_1$, the only
difference showing up in the $\Omega$ factors and
in the logarithm. The processes
with no identical particles in the final state are,
of course, easier to calculate.
They lead to limits on different combinations of $\Omega$ factors
but give no qualitatively new information.

%

\vspace{5mm}

We thank R.\ Holman, L.\ Lavoura, L.-F.\ Li and H.\ Vogel for
useful discussions.
We are also indebted to M.\ Savage and L.\ Wolfenstein
for useful suggestions and for reading and
criticizing the manuscript.
The work of J.\ P.\ S.\ was partially
supported by the Portuguese JNICT,
under CI\^{E}NCIA grant \# BD/374/90-RM,
and partially supported by
the United States Department of Energy,
under the contract \# DE-FG02-91ER-40682.
J.\ P.\ S.\ is also indebted to the Santa Barbara Institute for Theoretical
Physics where portions of this work were done.


%

\newpage


FIGURE CAPTIONS

\vspace{5mm}

Figure 1: One-loop contributions to the lepton flavour violating
$l_2 \rightarrow l_1 \gamma$ vertex.
The diagrams in figures 1b and 1c only contribute to the
vector and axial-vector parameters.

\vspace{5mm}

Figure 2: One-loop contributions to the process $l_2
\rightarrow 3 l_1$.
In our model, there are contributions from
a penguin diagram (fig.~2a) and from
a HH-box diagram (fig.~2b).


\begin{thebibliography}{99}
%
\bibitem{gsw}
S.\ L.\ Glashow,
Nucl.\ Phys.\  {\bf B22}, 579 (1961);\\
S.\ Weinberg,
Phys.\ Rev.\ Lett.\ {\bf 19}, 1264 (1967);\\
A.\ Salam,
in: Proceedings of the 8th Nobel Symposium,
ed. N.\ Svartholm,
(Almquist and Wiksell, Stockholm,1968), p.\ 367.
%
\bibitem{seesaw}
M.\ Gell-Mann, P.\ Ramond and R.\ Slansky,
in: Supergravity, ed. P.\ van Nieuwenhuizen and D.\ H.\ Freedman
(North-Holland, Amsterdam,1979).
%
\bibitem{nounit}
D.\ Wyler and L.\ Wolfenstein,
Nucl.\ Phys.\  {\bf B218}, 205 (1983);\\
J.\ Bernab\'{e}u, {\it et. al.},
Phys.\ Lett.\ B {\bf 187}, 303 (1987).
%
\bibitem{valle}
J.\ W.\ F.\ Valle,
Phys.\ Lett.\ B {\bf 199}, 432 (1987).
%
\bibitem{susy}
E.\ Roulet,
Phys.\ Rev.\ D {\bf 44}, R935 (1991)
and {\it ibid.} {\bf 44}, 3971 (1991);\\
M.\ M.\ Guzzo, A.\ Masiero and S.\ T.\ Petcov,
Phys.\ Lett.\ B {\bf 260}, 154 (1991);\\
V.\ Barger, R.\ J.\ N.\ Phillips and K.\ Whisnant,
Phys.\ Rev.\ D {\bf44}, 1629 (1991); \\
D.\ Kapetanakis, P.\ Mayr and H.\ P.\ Nilles,
Phys.\ Lett.\ B {\bf 282}, 95 (1992).
%
\bibitem{spcpv}
T.\ D.\ Lee,
Phys.\ Rev.\ D {\bf 8}, 1226 (1973). \\
See also S.\ Weinberg,
Phys.\ Rev.\ Lett.\ {\bf 37}, 657 (1976); \\
G.\ C.\ Branco,
Phys.\ Rev.\ Lett.\ {\bf 44}, 504 (1980).
%
\bibitem{nfc}
S.\ L.\ Glashow and S.\ Weinberg,
Phys.\ Rev.\ D {\bf 15}, 1958 (1977);\\
E.\ Paschos,
{\it ibid}.\ {\bf 15}, 1966 (1977).
%
\bibitem{pdb}
Particle Data Group,
K.\ Hikasa {\it et al.},
Phys.\ Rev.\ D {\bf 45}, 3262 (1992).
%
\bibitem{GWF}
C.\ Greub, D.\ Wyler and W.\ Fetscher,
Zurich University report 21/93, unpublished.
%
\bibitem{hayes}
See for example K.\ G.\ Hayes' review
in p. VI.19 of reference \cite{pdb}
and references therein.
%
\bibitem{erratum}
D.\ S.\ Akerib {\it et al},
Phys.\ Rev.\ Lett.\ {\bf 71}, 3395 (1993).
%
\bibitem{chengli}
See, for example, T.\ P.\ Cheng and L.-F.\ Li,
{\it Gauge Theories of Weak Interactions},
(Oxford University Press, 1985).
%
\bibitem{bp}
See, for example, S.\ M.\ Bilenky and B.\ Pontecorvo,
Phys.\ Rep.\ {\bf 41C}, 225 (1978).
%
\bibitem{gim}
S.\ L.\ Glashow, J.\ Iliopoulos and L.\ Maiani,
Phys. Rev. D {\bf 2}, 1285 (1970).
%
\bibitem{chengli2}
T.\ P.\ Cheng and L.-F.\ Li,
Phys. Rev. Lett. {\bf 45}, 1908 (1980);\\
also of interest are,
T.\ P.\ Cheng and L.-F.\ Li,
Phys. Rev. D {\bf 16}, 1425 (1977);\\
B.\ W.\ Lee and R.\ E.\ Schrock,
Phys. Rev. D {\bf 16}, 1444 (1977).
%
\bibitem{soares}
In performing the on-mass-shell renormalization we used
extensively the formalism set forth in
J.\ M.\ Soares and A.\ Barroso,
Phys. Rev. D {\bf 39}, 1973 (1989).
%
\end{thebibliography}
\end{document}